\documentclass{article}

\usepackage{amsmath,amssymb,upgreek}

\newcommand{\dd}{\mathrm{d}}
\newcommand{\ee}{\mathrm{e}}
\newcommand{\ii}{\mathrm{i}}

\newcommand{\C}{\mathbf{C}}

\newcommand{\D}{\mathrm{D}}

\renewcommand{\Re}{\mathrm{Re}\,}
\renewcommand{\Im}{\mathrm{Im}\,}

\begin{document}

\author{Andrzej Hanyga\\ ul. Bitwy Warszawskiej 14/52, 02-366 Warszawa PL\\
\tt{ajhbergen@yahoo.com}}

\title{Effects of Newtonian viscosity and relaxation on linear viscoelastic wave propagation}
\date{}

\maketitle

\begin{abstract}
In an important class of linear viscoelastic media the stress is the superposition of a Newtonian term and a stress relaxation term.
It is assumed that the creep compliance is a Bernstein class function, which entails that the relaxation function is LICM.
In this paper the effect of Newtonian viscosity term on wave propagation is
examined. It is shown that Newtonian viscosity dominates over the features resulting from stress relaxation. 
For comparison the effect of unbounded relaxation function is also examined. In both cases the wave propagation speed 
is infinite, but the high-frequency asymptotic behavior of attenuation is different.
Various combinations of Newtonian viscosity and relaxation functions and the corresponding creep compliances are summarized. 

\textbf{Keywords:} viscoelasticity, Newtonian viscosity, stress relaxation function, creep compliance,  LICM functions,
Bernstein functions

\textbf{MSC classfication:} MSC 74D05, MSC 74J05
\end{abstract}

\section{Introduction}

Creep behavior is the most direct result of experimental investigation of viscoelastic media. On the other hand initial and boundary-value 
problems for viscoelastic bodies involve Newtonian (instantaneous) viscosity and the stress relaxation function. It is therefore 
important  to connect the creep compliance with the Newtonian viscosity and stress relaxation at both the qualitative and quantitative
level. This subject is investigated in \cite{1}. 

In this paper we shall study how one-dimensional wave propagation characteristics 
are affected by creep, Newtonian viscosity and stress relaxation behavior. We investigate how these constitutive properties of 
the medium affect wave propagation speed and wave attenuation. In contrast to our earlier paper \cite{2} we shall take into account 
the Newtonian viscosity component, which was ignored in \cite{2}.

In this paper shall consider general viscoelastic media with a creep compliance in the Bernstein function class.
This class is sufficiently large to include Newtonian viscosity as well as completely monotone stress relaxation functions.
Furthermore, this class encompasses all the specific models of linear viscoelastic media in the literature. Although no convincing 
theoretical justification of this fact has ever been given it is reasonable to use the rich theory of Bernstein and LICM functions 
so long as the viscoelastic model under consideration belongs to this class.

In \cite{1} we showed that the viscoelastic stress relaxation corresponding to a general Bernstein class creep compliance \cite{2} 
in general involves a Newtonian viscosity term in addition to the stress relaxation term. The latter is assumed to be given by 
a Volterra convolution of the strain rate with a locally integrable completely monotone (LICM) kernel. More specifically, it will
be shown that a non-trivial Newtonian viscosity term
is present whenever the Bernstein class creep compliance does not involve an initial jump and has a finite initial slope. It is 
natural to expect the last condition to be usually satisfied. In other words, if the strain builds up gradually in response to a 
stress jump then stress-strain relation involves a Newtonian viscosity term in addition to a the stress relaxation term. We can thus 
infer about the presence or absence of the Newtonian viscosity term in the stress-strain relation from an examination of the results of 
creep tests.

In our earlier paper \cite{2} 
on linear dispersion and attenuation in linear viscoelastic media the Newtonian viscosity term was however ignored. We 
shall now examine the effect of including Newtonian viscosity term in the stress-strain relation in addition to the stress 
relaxation term. 
Many results obtained in \cite{2} still apply to the case considered here, hence we shall consider the present paper as 
a complement to \cite{2} and use the theorems proved in that paper while focusing on the differences between the linear viscoelastic 
models with and without the Newtonian viscosity term.

As expected, the main effect of Newtonian viscosity on wave propagation turns out to be infinite wave propagation speed. 
We shall show here that in the presence of a Newtonian viscosity term the high-frequency asymptotic attenuation function is proportional 
to $\omega^{1/2}$, where $\omega$ denotes the circular frequency, independently of the stress relaxation function.

Strong singularity of the relaxation function $G(t)$ at $t = 0$ (the case of unbounded $G(t)$) is another potential source of infinite wave propagation speed 
while Newtonian viscosity may be absent. This case was studied in Sec.~8 of \cite{2} (cf also \cite{11} in a more general context.).

Weak singularity of the relaxation function ($G_0 < \infty$, $G^\prime(t)$ is integrable on a neighborhood of 0) and absence of the Newtonian viscosity term entail finite propagation
speed and continuity of the wave field and its derivatives at the wave front (cf \cite{2,5,6,7,7a,7b,7c,7e,8} and, in a more general context, \cite{9,10,12}).
In the last papers it was not assumed that the relaxation function was completely monotone and, as a result, the relation to creep compliance was not considered. 

In media with a Newtonian viscosity component high-frequency wave attenuation is essentially determined by just 
the Newtonian viscosity coefficient. On the other hand the stress relaxation term is more relevant for low-frequency attenuation
and the results obtained in \cite{2} for low-frequency attenuation remain valid. Experimental results are as a rule to a low frequency
range.

\section{Relations between Newtonian viscosity, relaxation function and creep compliance.}

We shall consider the one-dimensional linear viscoelastic problem:
\begin{eqnarray} \label{diff}
\rho\, u_{,tt} & = & N\, u_{,txx} + G(t) \ast u_{,txx} + \delta(x), \; t \geq 0\\
u(0,x) & = &0, \\
u_{,t}(0,x) & =  & \delta(x)
\end{eqnarray}
where $u(t,x)$ denotes the viscoelastic displacement field, $N$ is the Newtonian viscosity coefficient, $N \geq 0$, and $G(t)$ 
is the stress relaxation function, assumed locally integrable and completely monotone (LICM) \cite{2}.

We recall that an infinitely differentiable real function $f(t)$ defined on the the set of non-negative reals is said to be completely monotone if $(-1)^n\, \D^n \, f(t) \geq 0$ for all $t > 0$ and for $n = 0,1, \ldots$, where $\D^n f$ denotes the $n$-th order derivative.
The function $f$ is continuous on the set of positive reals but it can tend to infinity at 0. It is therefore locally integrable if and
only if it is integrable over the set $0 \leq t \leq 1$.

The asterisk denotes the Volterra convolution 
\begin{equation}
(f \ast g)(t) := \int_0^\infty f(s)\, g(t - s) \, \dd s.
\end{equation}

Note that the second term on the right-hand side of \eqref{diff} includes elastic stress if 
$G(t) \geq E > 0$.

For $N = 0$ the above problem was investigated in much detail in \cite{2}. In this paper we shall focus here on the effect of $N > 0$.
For comparison we shall also examine the case of unbounded $G(t)$, because both cases result in infinite propagation speeds. In 
the summary we shall also include the weakly singular case ($G(0) < \infty$, $G^\prime(t)$ singular, but integrable over $[0,1]$)

In the problem under consideration the stress is given by the sum of the Newtonian term and the stress relaxation term:  
$\sigma = N\, u_{,tx} + G(t) \ast u_{,tx}$.
The strain can be expressed in terms of the stress by the formula $u_{,x} = C(t) \ast \sigma_{,t}$, where the function $C(t)$ is the 
creep compliance. Comparison of the stress-strain and strain-stress constitutive equations yields the well-known duality relation 
\begin{equation} \label{duality}
N \, C(t) + G(t) \ast C(t) = t
\end{equation}
\cite{2}. Applying the Laplace transformation 
we get the equivalent relation
\begin{equation} \label{Lapdual}
\left[ N\, p + p\, \tilde{G}(p) \right] \, p\, \tilde{C}(p) = 1
\end{equation}  
where
\begin{equation}
\tilde{f}(p) := \int_0^\infty \ee^{-p t} \, f(t) \, \dd t
\end{equation}

The creep compliance $C(t)$ is a continuous, non-decreasing and non-negative function, hence it has a finite non-negative limit 
at 0, which we denote by $C(0)$.

We shall assume here that the creep compliance $C(t)$ is a Bernstein function, that is $C(t) \geq 0$ and 
$(-1)^n\, \D^n \, C(t) \leq 0$ for $t > 0$ and $n = 1, 2,\ \ldots$.

We know \cite{1} that for a Bernstein function $C(t)$ the solution of the duality relation has the form $N\, u(t) + G(t)$,
where $N$ is a non-negative number, $G(t)$ is a LICM function and $u(t)$ denotes the unit in the convolution algebra of 
functions\footnote{$u$ is not a function and thus it is not an element of the convolution algebra. This problem can be easily 
circumvented by using the concept of an approximate unit.}
on $[0,\infty[$. 

The derivative $C^\prime(t)$ of the creep compliance $C(t)$ is also a LICM function, hence it is non-negative and non-increasing.  
It follows that $C^\prime(0) \geq 0$, but it can be infinite.

If $G$ is bounded then $G_0 := G(0) > 0$, because $G$ is non-negative non-increasing and not identically 0. $G(t)$ is however often
unbounded at $t = 0$.
We shall define $G_0 = \infty$ if $G$ is unbounded.

In view of the identity $p\,\tilde{C}(p) = \widetilde{C^\prime}(p) + C(0)$ equation~\eqref{Lapdual} implies that
$$N \, p\, \widetilde{C^\prime}(p) + N\,p\, C(0)  + p\, \tilde{G}(p)\, p\, \tilde{C}(p) = 1$$
with $N, C(0), \widetilde{C^\prime}(p), \tilde{G}(p), \tilde{C}(p) \geq 0$. 

If $N > 0$ then in the limit $p \rightarrow \infty$ we get the 
equations $C(0) = 0$ and 
$N\, C^\prime(0) + G_0\, C(0) = 1$ (see Appendix for $G_0 = \infty$). Thus $N > 0$ implies that $C(0) = 0$ and $C^\prime(0) < \infty$. 
Hence if $N > 0$ and $G_0 < \infty$, then
\begin{equation}  \label{curious}
N \, C^\prime(0) = 1
\end{equation}

Furthermore $G_0 = \infty$ implies that $C(0) = 0$.

Equation~\eqref{curious} allows an estimate of the Newtonian viscosity coefficient from creep data.

It is worth noting that $C^\prime(0) = 0$ implies that $C^\prime(t) = 0$ for all $t \geq 0$ (because $C^\prime$ is non-negative and 
non-increasing), 
hence $C(t) = a$ for some non-negative constant $a$ and for all $t \geq 0$; by the duality relation \eqref{duality} 
$a \int_0^t G(s)\ \dd s = t - a\, N$. Hence
$a\, G(t) = 1$ for $t \geq 0$ and $a > 0$ and therefore $N = 0$. Hence in this case 
we are dealing with pure elastic stress. Hence $C^\prime(0) > 0$ and $N = 1/C^\prime(0)$ if the pure elastic case is excluded.

In the alternative case the Newtonian viscosity coefficient $N$ is either 0 or it is 
determined by equation~\eqref{curious}.

\section{Wave attenuation in linear viscoelastic media with Newtonian viscosity.}

Upon applying Laplace and Fourier transformation the wave equation assumes the form
\begin{equation} \label{wvf}
\rho \, p^2\, U(p,k) = -\left[ N\, p \, k^2 + p\, k^2 \, \tilde{G}(p) \right]\,  U(p,k) + 1
\end{equation}
here $U(p,k)$ denotes the simultaneous Laplace transform of $u(t,x)$ with respect to $t$ and the Fourier transform with respect to $x$.
The dispersion equation for equation~\eqref{wvf} is 
\begin{equation} \label{disp}
\rho\, p^2 + \left[ N \, p + p \, \tilde{G}(p) \right] \, k^2 = 0
\end{equation}

We shall change the variable $k$:
$ k = - \ii \kappa(p)$. The solution of equation~\eqref{disp}
can be expressed in the following form
\begin{equation} \label{kappa}
\kappa(p)/p = \rho^{1/2} / \left[ N\, p + p \, \tilde{G}(p)\right]^{1/2}
\end{equation}
with the square root chosen so that $\Re \kappa(p) \geq 0$ for $\Re p > 0$. The function $\kappa(p)$ is known as the 
complex wavenumber function \cite{2}.

The wave field is given by the inverse Fourier and Laplace transform of $U(p,k)$. Upon working out the inverse Fourier transform 
we end up with the following expression \cite{2}:
\begin{equation} \label{wavef}
u(t,x) = \frac{\rho}{4 \uppi \ii} \int_{-\ii \infty + \varepsilon}^{\ii \infty + \varepsilon} 
\frac{1}{\left[N\, p + p\, \tilde{G}(p)\right]\, \kappa(p)}\,
\ee^{p\, t - \kappa(p) \, \vert x \vert} \, \dd p
\end{equation}

Since $G$ is LICM, Bernstein's Theorem \cite{2,4} implies that $G(t)$ is the Laplace transform of a Borel measure $\mu$ on
the set $[0,\infty[$ of non-negative real numbers 
\begin{equation}
G(t) = \int_{[0,\infty[} \ee^{-r\, t}\, \mu(\dd r).
\end{equation}
The function $G(t)$ is locally integrable or, equivalently, integrable over the interval $0 \leq t \leq 1$ if and only if 
\begin{equation} \label{LICM}
\int_{[0,\infty[} \frac{1}{1 + r}\, \mu(\dd r) < \infty
\end{equation}
\cite{2}. 

For $p > 1$ we have 
$$\tilde{G}(p) = \int_{[0,\infty[} \frac{1}{p + r} \,\mu(\dd r) 
\leq \int_{[0,\infty[} \frac{1}{1 + r}\, \mu(\dd r) < \infty.$$
By the Lebesgue Dominated Convergence Theorem 
\begin{equation} \label{dd}
\lim_{p\rightarrow \infty} \tilde{G}(p) = 0
\end{equation} 
and therefore the term $N\, p$ in the denominator of equation~\eqref{kappa} dominates for large $p$. This holds for finite and infinite $G_0$.

The limit of $\kappa(p)/p$  at $p \rightarrow \infty$ equals
\begin{equation}
\lim_{p\rightarrow \infty, \; \Re p \geq 0} \kappa(p)/p = \left\{ \begin{array}{ll} 0, & \mbox{ $N > 0$ \, or\, $G_0 = \infty$}\\
1/c_\infty, & \mbox{ $N = 0$ \, and \, $G_0 < \infty$}
\end{array}
\right.
\end{equation}
where $c_\infty := \left[ G_0/\rho\right]^{1/2}$ if $G_0 < \infty$. 

It follows that for $N > 0$ the exponent in equation~\eqref{wavef} assumes the form  $p\, t - \mathrm{o}[p] \vert x \vert$.
For $t < 0$ the contour integral \eqref{wavef} can therefore be closed by a large half-circle in the right half of the $p$-plane,
where the integrand does not have any singularities. The wave field thus vanishes for $t < 0$, but in does not
vanish anywhere in the space
for $t > 0$. As expected, the disturbance spreads immediately to the entire space and the problem is
no longer hyperbolic.

We recall \cite{4,2} that a function $f(p)$ is a complete Bernstein function (CBF) if and only if it has the form $$f(p) = 
A\, p + p \, \int_{[0,\infty[} \frac{\mu(\dd r)}{p + r} $$
where $A \geq 0 $ and $\mu$ is a Borel measure on $[0,\infty[$ satisfying inequality \eqref{LICM}.
Therefore the function 
$N \, p + p \, \tilde{G}(p)$ is a CBF, hence by Theorem~2.7 in \cite{2} its square root is a CBF. By Theorem~2.8 \emph{ibidem} 
$\kappa(p)$ is a CBF. Hence $\kappa(p) = B\, p + \beta(p)$, where 
\begin{equation}
\beta(p) = p \int_{[0,\infty[} \frac{\nu(\dd r)}{r + p} = \mathrm{o}[p],
\end{equation}
$\nu$ is a Borel measure on $[0,\infty[$ satisfying the inequality
\begin{equation}
\int_{[0,\infty[} \frac{\nu(\dd r)}{r + 1} < \infty.
\end{equation}

Again we have that $\beta(p) = o[p]$ for $p \rightarrow \infty$, $\Re p > 0$, hence 
$$B = \lim_{p \rightarrow \infty, \, \Re p > 0} \kappa(p)/p,$$
so that $B = 1/c_\infty$  
if $N = 0$ and $G(t)$ is bounded, while $B = 0$ otherwise.

In the case of $N = 0$ the asymptotic behavior of $\kappa(p)$ for $p \rightarrow \infty$ was studied in \cite{2}.

For $N > 0$ we note that 
\begin{equation}
\lim_{p \rightarrow \infty,\, \Re p \geq 0} \kappa(p)/p^{1/2} = \left[N/\rho\right]^{-1/2}
\end{equation}
because of \eqref{dd}. It follows that 
for $N > 0$ 
\begin{equation}
\kappa(p) =  \left[N/\rho\right]^{-1/2} \, p^{1/2} + \mathrm{o}\left[ p^{1/2}\right]
\end{equation}
The high-frequency asymptotics of the attenuation function $a(\omega)$ for $N > 0$ is 
therefore given by the formula 
\begin{equation} \label{asatt}
a(\omega) := \Re \kappa(-\ii \omega) \sim_{\omega \rightarrow \infty} \left[N/\rho\right]^{-1/2} \, \vert \omega \vert^{1/2}/ \sqrt{2}
\end{equation}
The high-frequency asymptotic attenuation is thus entirely controlled by the Newtonian viscosity coefficient, with the relaxation 
term playing only a secondary role.

For comparison with \cite{2} we shall also note that by Valiron's theorem (\cite{2}, Thm B.4)  $\nu([0,r]) \sim_{r \rightarrow \infty} 
r^{1/2} \, l(r)$, where $l$ is a function slowly varying at infinity \cite{3}.

\section{The effect of the strong singularity of the relaxation function on the complex wave number function and attenuation.}
\label{singularity}

Another case of infinite wave propagation speed involves an unbounded $G(t)$. This case is referred to in \cite{2} as strong singularity.

We assume accordingly that $G(t) = t^{-\alpha}\, l(t)/\Gamma(1 - \alpha)$,
 where $0 < \alpha < 1$ and the function $l(t)$ is slowly varying at 0.  Excluding the case of $G(t) \equiv 0$, $G(t)$ is positive for sufficiently small 
 $t > 0$, hence  $l(t) > 0$. 

By the Karamata theorems \cite{3} $p\, \tilde{G}(p) = p^\alpha\, l(1/p)$, hence $p\, \tilde{C}(p) = p^{-\alpha}/l(1/p)$ and 
$C(0) = \lim_{p\rightarrow\infty} \left[p \, \tilde{C}(p)\right] = 0$. Furthermore $C^\prime(0) = \lim_{p\rightarrow \infty} 
\left[p^2\, \tilde{C}(p)\right] = \infty$.

The wave propagation parameters for this case are studied in Theorem~8.1 in \cite{2}. 
It is proved in Sec.~8 of \cite{2}, that in the case of $N = 0$ and $G$ strongly singular the propagation speed is infinite and 
the asymptotic attenuation $a(\omega) = l(\omega)\,  \omega^\gamma$ with $1/2 < \gamma < 1$, where $l(\omega)$ is slowly varying at 
infinity (Theorem~8.1). 

Consequently, an infinite propagation speed and an asymptotic attenuation $\propto \omega^\gamma$ with $1/2 < \gamma < 1$ indicates
that $N = 0$ and the stress relaxation function is singular, while an attenuation function $\propto \omega^{1/2}$ indicates that 
$N > 0$. Of course it may be difficult to distinguish between the two cases if $\gamma$ is close to $1/2$.

\section{A few examples}

We shall consider the creep compliances of three simple viscoelastic media with a Newtonian component and a stress relaxation. 
The relaxation function can be bounded or unbounded.\\

\textbf{Example 1.}
\textbf{Strongly singular relaxation function: $G_0 = \infty$.} \\

We assume the constitutive equation
\begin{equation}
\sigma = N \, \dot{e} + t^{-\alpha}\ast \dot{e}/\Gamma(1 - \alpha) =: R\ast \dot{e}, \;\; 0 < \alpha < 1,
\end{equation}
where $R = N\, \delta + t^{-\alpha}/\Gamma(1 - \alpha)$, $\tilde{R}(p) = N + p^{-1+\alpha}$. Hence 
$$C(0) = \lim_{p \rightarrow \infty} \left[p \, \tilde{C}(p)\right] = \lim_{p\rightarrow \infty} \,\left[N \, p + p^\alpha\,\right]^{-1} = 0$$

Furthermore 
$$C^\prime(t) = \frac{1}{2 \uppi \ii} \int_{-\ii \infty}^{\ii \infty} \ee^{p t} \, \left[ N \, p + p^\alpha \right]^{-1} \, \dd p$$
where the Bromwich contour $\mathcal{B}$ runs along the imaginary axis to the right of it.
We now check whether the integrand has poles 
\begin{equation}\label{z}
N\, p + p^\alpha = 0.
\end{equation}
 At a pole $p = R \, \ee^{\ii\, \phi}$ we must have  
$N \, R \, \sin(\phi) + R^\alpha \, \sin(\alpha\, \phi) = 0$. For $0 < \phi \leq \uppi$ both  terms are positive, hence 
equation~\eqref{z} cannot be satisfied. However for $\phi = 0$ the real part of \eqref{z} $N \, R  + R^\alpha$ can vanish only 
for $R = 0$, $p = 0$. 

Since the integrand vanishes exponentially for $\Re p < 0$ and the only singularity is the branching cut $]-\infty,0]$, 
the Bromwich contour can be deformed into a contour running from $-\infty$ to 0 below the negative real half-axis  and then above 
the negative real half-axis from 0 to $-\infty$ with a half-circle $K$ of a very small radius $\varepsilon$ in the right-half plane. The contribution  
$\ii p \int_{-\uppi/2}^{\uppi/2} \ee^{t p} \dd \phi/\left[N p + p^\alpha\right]$ with $p = \varepsilon \, \ee^{\ii \phi}$ 
of the half-circle $K$
vanishes in the limit $\varepsilon \rightarrow 0$, hence we obtain the formula
\begin{eqnarray}\label{pu}
C^\prime(t) = \frac{1}{\uppi} \Im \int_{-\infty}^0 \ee^{-r t} \, \frac{1}{-N r + r^\alpha\, \ee^{-\ii \uppi \alpha}} \, \dd r = \\
\frac{\sin(\alpha \uppi}{\uppi} \int_0^\infty \ee^{-r t}\, \frac{r^\alpha\, \dd r}{(-N \,r + r^\alpha \, \cos(\alpha \uppi))^2 + r^{2 \alpha}\, \sin^2(\alpha \uppi)} 
\end{eqnarray}
For $t = 0$ the integrand of the integral on the right-hand side is integrable if and only if $N > 0$, 
thus $C^\prime(0) < \infty$ if $N > 0$ and $C^\prime(0) = \infty$ otherwise.\\

\textbf{Example 2.}
\textbf{$G_0 < \infty$, $N > 0$.}\\

For example for 
$$\sigma = N \, \dot{e} + \ee^{-t}\ast \dot{e}.$$
we have $\tilde{R}(p) = N + 1/(p + 1)$. Hence 
\begin{equation*}
C(0) = \left\{ \begin{array}{ll} 0 & \textrm{if} \, N > 0 \\
1 & \textrm{if} \, N = 0 \end{array} \right\} 
\end{equation*}
and
$$C^\prime(t) = \frac{1}{2 \uppi \ii} \int_{-\ii \infty}^{\ii \infty} \frac{\ee^{p t}}{p} \frac{1}{N\, p + N + 1}\, \dd p.$$
The the integrand in the last formula does not have a branching cut but it has two residues at $p = 0$ 
and (if $N > 0$) also $p = -P$, where $P := (N +1)/N$. For $N > 0$ we thus have
$C^\prime(t) = \left[ 1 + \ee^{-P t}\right]/(N + 1)$ and $C^\prime(0) = 2/(N + 1) \, \dd p$. 
Consequently if $N > 0$, then $C(0) = 0$,  while $C^\prime(0)$ is finite.

As $N \rightarrow 0$ the parameter $P \rightarrow \infty$ and $C^\prime(t)$ tends to a constant. Thus for $N = 0$ we have 
$C(0) > 0, C^\prime(0) < \infty$. \\

\textbf{Example 3.}
\textbf{Weakly singular relaxation function: $G_0 < \infty$, $G^\prime$ has a locally integrable singularity.}\\

In the weakly singular case the relaxation function is bounded, but its derivative has a locally integrable singularity \cite{2}.
Here is an example of a weakly singular relaxation function, known as the KWW or stretched exponential model in viscoelasticity and optics:
$$G(t) = \ee^{-t^\alpha}, \; \; 0 < \alpha < 1$$

The function $G(t)$ is CM because it is a superposition of a CM function $\ee^{-x}$ and a Bernstein function 
$x = t^\alpha$, $0 < \alpha < 1$.

$$G^\prime(t) = -\alpha\, t^{\alpha-1} \, \ee^{-t^\alpha}$$
$$p\, \tilde{G}(p) = \left[ 1 - p^{-1/2}/2 \sqrt{\uppi}\right] \, \ee^{1/(4 p)} \rightarrow_\infty G_0 = 1 $$
$$p\, \tilde{C}(p) = \left[ 1 - p^{-1/2}/2 \sqrt{\uppi}\right]^{-1} \, \ee^{-1/(4 p)} \rightarrow_\infty C(0) = 1 $$
$$ \lim_{t\rightarrow 0} C^\prime(t) = \lim_{p\rightarrow\infty} p\, \left[p \, \tilde{C}(p) - C(0)\right] = \infty $$

\section{Conclusions}

For a general creep compliance in the Bernstein function class stress can be expressed in terms of strain rate as a superposition 
of a Newtonian viscosity term and a stress relaxation term with a LICM relaxation function.

Linear viscoelastic media with a Bernstein class creep compliance can be divided into two categories: (1) media with an initial 
jump of the creep compliance ($C(0) > 0$) or with no jump but $C^\prime(0) = \infty$;
(2) media with no initial jump of creep compliance ($C(0) = 0$) and with finite $C^\prime(0)$. In the second class the stress 
always contains a Newtonian term, while in the first class there is no Newtonian stress component. In this class there are also 
media with an unbounded LICM relaxation function.

If the Newtonian viscosity coefficient is positive or the relaxation function is unbounded, then the speed of propagation is infinite.
In the first case the high-frequency asymptotic attenuation is essentially determined by the Newtonian viscosity coefficient, 
while the stress relaxation 
term only influences less important corrections to the attenuation. In the second case the asymptotics of the 
attenuation is different so that the two cases can in principle be distinguished by examining the asymptotics of the attenuation.

Low-frequency asymptotics of $\kappa(p)/p$ is controlled by the low-frequency asymptotics of the function $p\, \tilde{G}(p)$, while 
the term $N p$ in the denominator of~\eqref{kappa} plays a secondary role. Hence for low frequencies 
the results of \cite{2} still apply.

The relation between creep, relaxation, wave speed and the attenuation function are summarized in the table below.\\ \vspace{0.2cm}

\begin{table}
\begin{tabular}{|| l | l | l ||}
\hline\hline
Creep parameters & Relaxation parameters & Wave propagation parameters \\
\hline\hline
$C(0) > 0$ &  $N = 0$, $G_0 < \infty$, $G^\prime(t) \sim_0 -A \, t^{-\alpha}$, 
$0 < \alpha < 1$ & $c_\infty < \infty$, $a(\omega) \propto \omega^\beta$, $0 < \beta < 1$  \\
\hline
$C(0) = 0$, $\C^\prime(0) = \infty$ & $N = 0$, $G(t) \sim_0 A\, t^{-\alpha}$, $0 < \alpha < 1$ & $c_\infty = \infty$, $a(\omega) \propto \omega^\gamma$, $1/2 < \gamma < 1$\\
\hline
$C(0) = 0$, $C^\prime(0) < \infty$ & $N > 0$ & $c_\omega = \infty$, $a(\omega) \propto \omega^{1/2}$\\
\hline\hline
\end{tabular}\\ \vspace{0.2cm}
\caption{Summary of basic relationships between creep compliance, stress relaxation and wave propagation.
$\beta = 1 - \alpha$, $\gamma = 1 - \alpha/2$.  } \label{t1}
\end{table}

Table~\ref{t1} shows that the coefficient of the Newtonian viscosity as well as some parameters of the stress relaxation and 
wave attenuation can be estimated from creep data.

The first row in Table~\ref{t1} represents weakly singular relaxation \cite{2}. In this case $G_0 < \infty$ and  $G^\prime$ is 
assumed to have a locally 
integrable singularity. In this case the boundary of the region perturbed by the waves does not involve
any discontinuity of the field nor of any of its time and space derivatives. The last statement follows from the fact that the factor 
$\ee^{-a(\omega)\, \vert x \vert}$ ensures absolute convergence with respect of $\omega$ in the Fourier transforms of the Green's function
and its derivatives \cite{2}. This case is most important in applied viscoelasticity and in poroelasticity \cite{5,6,7a,7b,7c,7e,8,5a} as it involves finite 
propagation speed of the wavefront and a delay of the first pulse arrival with respect to the wavefront. The delay should be taken 
into account in inverse methods for the a correct determination of scatterer positions \cite{5}. 

Note that in all the cases the attenuation function is sublinear. This follows from concavity of the creep compliance \cite{7d} and 
also from causality \cite{7c}. This fact is worth mentioning because its general validity has sometimes been questioned.

The second row represents the strongly singular relaxation and the third row represents the effects of nontrivial Newtonian viscosity.

\section*{Appendix}

\appendix 

It is well-known that
\begin{equation} \label{xxx} 
G(0) = \lim_{p\rightarrow\infty} \left[ p \, \tilde{G}(p) \right]
\end{equation}
if $G(0)$ is finite. We shall now extend this equation to unbounded completely monotone functions.

Using Bernstein's theorem \cite{4},
$$ p \, \tilde{G}(p) = p \int_{[0,\infty[} \frac{1}{p + r} \, \mu(\dd r)$$
hence $p \, \tilde{G}(p)$ is a non-decreasing function. If it is bounded by a number $C$, then 
$$ G(t) = \int_{[0,\infty[} \ee^{-r t} \, \mu(\dd r) \leq 1/t \int_{[0,\infty[} \frac{1}{1/t + r} \, \mu(\dd r) \leq C$$
for $t > 0$, hence $G_0 < \infty$. 

If $G_0 = \infty$, then $p \, \tilde{G}(p)$ is non-decreasing and unbounded, hence it tends to infinity for $p \rightarrow \infty$.\\
\mbox{ }\hfill$\Box$

\section*{ Note.}

This article has appeared  in Archive of Applied Mechanics, doi:10.1007/s00419-019-01620-2.

\end{document}